\def\lesssim{{_ <\atop{^\sim}}}

\def\ap3m{AP$^3$M}
\def\LCDM{$\Lambda$CDM}

\def\hkpc{$h^{-1}{\ }{\rm kpc}$}
\def\hMpc{$h^{-1}{\ }{\rm Mpc}$}
\def\hMsun{$h^{-1}{\ }{\rm M_{\odot}}$}

\def\nbody{$N$-body}
\def\c15{$c_{\rm 1/5}$}

\def\vecx{\vec{x}}
\def\vecp{\vec{p}}
\def\vecr{\vec{r}}
\def\vecF{\vec{F}}
\def\vecg{\vec{g}}

\def\vecnabla{\vec{\nabla}}
\def\dvecx{\dot{\vec{x}}}
\def\ddvecx{\ddot{\vec{x}}}
\def\ddvecr{\ddot{\vec{r}}}

\newcommand{\Table}[1]{Table~\ref{#1}}

\newcommand{\Eq}[1]{Eq.~(\ref{#1})}
\newcommand{\Fig}[1]{Fig.~\ref{#1}}
\newcommand{\mlapm}{\texttt{MLAPM}}

\def\ea{et~al.~}                            % \ea      =  et al.
\def\lesssim{\mathrel{\hbox{\rlap{\hbox{\lower4pt\hbox{$\sim$}}}\hbox{$<$}}}}
\def\gtrsim{\mathrel{\hbox{\rlap{\hbox{\lower4pt\hbox{$\sim$}}}\hbox{$>$}}}}

\newcommand{\AAA}[3]    {\mbox{A\&A~\textbf{#1},~#2~(#3)}}

\newcommand{\ApJ}[3]    {\mbox{ApJ~\textbf{#1},~#2~(#3)}}
\newcommand{\ApJS}[3]   {\mbox{ApJ~Suppl.~\textbf{#1},~#2~(#3)}}
\newcommand{\ApJL}[3]   {\mbox{ApJ~Lett.~\textbf{#1},~#2~(#3)}}
\newcommand{\ARAA}[3]   {\mbox{Ann.~Rev.~A~\&~A~\textbf{#1},~#2~(#3)}}
\newcommand{\AJ}[3]     {\mbox{Astron.~J.~\textbf{#1},~#2~(#3)}}
\newcommand{\MNRAS}[3]  {\mbox{MNRAS~\textbf{#1},~#2~(#3)}}

\newcommand{\astroph}[1]{\mbox{\texttt{astro-ph/#1}}}

\documentstyle[epsfig]{mn}

\begin{document}

%%%%%%%%%%%%%%%%%%%%%%%%%%%%%%%%%%%%%%%%%%%%%%%%%%%%%%%%%%%%%%%%%%%%%%%%
%                               TITLE                                  %
%%%%%%%%%%%%%%%%%%%%%%%%%%%%%%%%%%%%%%%%%%%%%%%%%%%%%%%%%%%%%%%%%%%%%%%%
\title{Galactic Halos in MONDian Cosmological Simulations}

\author[Knebe A. \& Gibson B.K.]
       {Alexander Knebe and Brad K. Gibson\\        
       {Centre for Astrophysics \& Supercomputing,
        Swinburne University, P.O. Box 218, Mail \# 31,
        Hawthorn, Victoria, 3122, Australia}}

\date{Received ...; accepted ...}

\maketitle

%%%%%%%%%%%%%%%%%%%%%%%%%%%%%%%%%%%%%%%%%%%%%%%%%%%%%%%%%%%%%%%%%%%%%%%%
%                             ABSTRACT                                 %
%%%%%%%%%%%%%%%%%%%%%%%%%%%%%%%%%%%%%%%%%%%%%%%%%%%%%%%%%%%%%%%%%%%%%%%%
\begin{abstract}
In this paper a series of high-resolution \nbody\ simulations is
presented in which the equations of motion have been changed to
account for MOdified Newtonian Dynamics (MOND). It is shown that a
low-$\Omega_0$ MONDian model with an appropriate choice for the
normalisation $\sigma_8$ can lead to similar clustering properties at
redshift $z=0$ as the commonly accepted (standard) \LCDM\ model.
However, such a model shows no significant structures at high redshift
with only very few objects present beyond $z>3$ that can be readily
ascribed to the low $\Omega_0$ value adopted. The agreement with
\LCDM\ at redshift $z=0$ is driven by the more rapid structure
evolution in MOND. Moreover, galaxy formation appears to be more
strongly biased in MONDian cosmologies. Within the current
implementation of MOND density profiles of gravitationally bound
objects at $z=0$ can still be fitted by the universal NFW profile but
MOND halos are less clumpy.
\end{abstract}


%%%%%%%%%%%%%%%%%%%%%%%%%%%%%%%%%%%%%%%%%%%%%%%%%%%%%%%%%%%%%%%%%%%%%%%%
%                           INTRODUCTION                               %
%%%%%%%%%%%%%%%%%%%%%%%%%%%%%%%%%%%%%%%%%%%%%%%%%%%%%%%%%%%%%%%%%%%%%%%%
\section{Introduction}
Although the currently favoured \LCDM\ model has proven to be
remarkably successful on large scales (cf. WMAP results, Spergel et
al. 2003), recent high-resolution \nbody\ simulations seem to be in
contradiction with observation on sub-galactic scales: the Cold Dark
Matter "crisis" on small scales is far from being over. The problem
with the steep central densities of galactic halos, for instance, is
still unsolved as the highest resolution simulations favor a cusp with
a logarithmic inner slope for the density profile of approximately
-1.2 (Power~\ea 2003), whereas high resolution observations of low
surface brightness galaxies are best fit by halos with a core of
constant density (de Block \& Bosma 2002; Swaters~\ea 2003). Suggested
solutions to this include the introduction of self-interactions into
collisionless \nbody\ simulations (e.g. Spergel~\& Steinhardt 2000;
Bento~\ea 2000), replacing cold dark matter with warm dark matter
(e.g. Knebe~\ea 2002; Bode, Ostriker~\& Turok 2001; Colin~\ea 2000) or
non-standard modifications to an otherwise unperturbed CDM power
spectrum (e.g. bumpy power spectra; Little, Knebe~\& Islam 2003;
tilted CDM; Bullock 2001c). Some of the problems, as for instance the
overabundance of satellites, can be resolved with such modifications
but none of the proposed solutions have been able to rectify
\textit{all} shortcomings of \LCDM\ simultaneously.

Therefore, there might be alternative solutions worthy of exploration,
one of which is to abandon dark matter completely and to adopt the
equations of MOdified Newtonian Dynamics (MOND; Milgrom 1983;
Bekenstein~\& Milgrom 1984). It has already been shown by other
authors that this simple idea might explain many properties of
galaxies without the need of non-baryonic matter (e.g. Scarpa 2003;
McGaugh~\& de Blok 1998; Sanders 1996; Milgrom 1994; Begeman,
Broeils~\& Sanders~\ea 1991). MOND is also successful in describing
the dynamics of galaxy groups and clusters (Sanders 1999; Milgrom
1998), globular clusters (Scarpa 2002) and, to a limited extent,
gravitational lensing (Mortlock~\& Turner 2001; Qin~\& Zou 1995). A
recent review of MOND is given by Sanders~\& McGaugh (2002) which also
summarizes (most of) the successes alluded to above.

However, there has yet to come a detailed study of the implications of
MOND in cosmological simulations of structure and galaxy formation,
which is the aim of the current study. Nusser (2002) already
investigated modified Newtonian dynamics of the large-scale structure
using the
\nbody\ approach. His simulations, however, are lower resolution,
both in terms of spatial and mass resolution, making a study of
individual objects difficult. Moreover, his implementation of the MOND
equations is slightly different to our treatment.

The outline of the paper is as follows. In Section~\ref{EOM} we
present the way we modified our \nbody\ code \mlapm\ to account for
MOND. Section~\ref{Models} introduces the cosmological models under
investigation whereas Section~\ref{Nbody} summarizes the numerical
details.  The analysis can be found in Section~\ref{Analysis}. We
finish with a summary and our conclusions in
Section~\ref{Conclusions}.

%%%%%%%%%%%%%%%%%%%%%%%%%%%%%%%%%%%%%%%%%%%%%%%%%%%%%%%%%%%%%%%%%%%%%%%%
%                         EQUATIONS OF MOTION                          %
%%%%%%%%%%%%%%%%%%%%%%%%%%%%%%%%%%%%%%%%%%%%%%%%%%%%%%%%%%%%%%%%%%%%%%%%
\section{The MONDian Equations of Motion} \label{EOM}

In an \nbody\ code one integrates the (comoving) equations of motion

\begin{equation}\label{eomPec}
 \begin{array}{rcl}
  \displaystyle \dot{\vecx}          
   & = &\displaystyle  {\vecp \over a^2} \ , \\ 
\\
  \displaystyle \dot{\vecp}          
   & = &\displaystyle  {\vecF_{\rm pec}\over a} \\
 \end{array}
\end{equation}

\noindent
which are completed by Poisson's equation 

\begin{equation}\label{PoissonPec}
  \displaystyle \vecnabla_x \cdot \vecF_{\rm pec}(\vecx) 
    = -\Delta_x \Phi(\vecx) = - 4 \pi G (\rho(\vecx) - \overline{\rho}) \ .
\end{equation}

In these equations $\vecx = \vecr/a$ is the comoving coordinate,
$\vecp$ the canonical momentum, $\vecnabla_x \cdot$ the divergence
operator ($\Delta_x$ the Nabla operator) with respect to $\vecx$ and
$\vecF_{\rm pec}(\vecx) = -\nabla
\Phi(\vecx)$ the \textit{peculiar acceleration field in comoving coordinates}.
We now need to to modify these (comoving) equations to account for
MOND.

Despite MOND being a modification to Newton's second law rather than
to gravity, one has the option to actually interpret MOND as an
alteration of the law of gravity (cf. Sanders~\& McGaugh 2002). In
that case Poisson's equation

\begin{equation} \label{NewtonPoisson}
 \vecnabla_r \cdot \vecg = -4\pi G \rho(\vecr)
\end{equation}

\noindent
is replaced by

\begin{equation} \label{MONDPoisson}
 \vecnabla_r \cdot \left(\mu \left({|\vecg_M|/g_0} \right) \vecg_M \right) 
  = - 4 \pi G \rho(\vecr) \ ,
\end{equation}

\noindent
where $\vecr = a \vecx$ is the proper coordinate, $g_0$ the
fundamental acceleration of MOND and $\vecg_M$ the MONDian
acceleration field. Note that \Eq{NewtonPoisson} and \Eq{MONDPoisson}
are given in proper coordinates in contrast to \Eq{PoissonPec} where
the solution $\vecF_{\rm pec}$ describes only the \textit{peculiar}
acceleration.

If we now compare \Eq{NewtonPoisson} and \Eq{MONDPoisson} we find the
relation between MONDian acceleration $g_M$ and Newtonian acceleration
$g$ to be:

\begin{equation} \label{curl}
 \vecg = \mu(\vecg_M/g_0) \vecg_M + \nabla \times \vec{h} \ .
\end{equation}

\noindent
The field equation~(\ref{MONDPoisson}) is non-linear and difficult to
solve in general. But it has been shown by Bekenstein~\& Milgrom
(1984) that $\nabla \times \vec{h}$ decreases like 
\textbf{\textit{O}}$(r^{-3})$ with increasing scale.
Moreover, in cases of spherical, planar, or cylindrical symmetry the
curl-term vanishes. Under the assumption that objects forming in the
Universe show at least one of these symmetries we are able to neglect
the curl-term completely. One might argue that this kind of symmetry
is probably very weak at high redshifts. However, later in this
Section we are also making the assumption that MOND only affects
peculiar accelerations (in proper coordinates) which are well above
the MOND acceleration at early times (cf. \Eq{PecAcc2}).  We therefore
presume that $\nabla \times \vec{h}$ is unimportant for the growth of
large-scale structures as well as for the internal properties of
virialized objects (under the assumption that they are at least
symmetric along one of their axes).

Using Milgrom's (1983) suggested interpolation function

\begin{equation} \label{MilgromInterpol}
 \mu(x) = x (1+x^2)^{-1/2}
\end{equation}

\noindent
one now only needs to solve

\begin{equation} \label{ggNsolve}
 g_M^2 - g \sqrt{g_0^2 + g_M^2} = 0 \ ,
\end{equation}

\noindent
to get $g_M$ as a function of $g$. The relevant solution of \Eq{ggNsolve} is

\begin{equation} \label{ggN}
 g_M = g \left(\frac{1}{2} + \frac{1}{2} \sqrt{1+\left(\frac{2 g_0}{g}\right)^2}\right)^{1/2} \ .
\end{equation}

\Eq{ggN} allows us to obtain the MONDian acceleration $\vecg_M$ for a given
Newtonian acceleration $\vecg$ where $\vecg_M$ and $\vecg$ are assumed
to be parallel.

However, we are actually solving \Eq{PoissonPec} in our \nbody\ code
\mlapm\ which gives $\vecF_{\rm pec}$, the Newtonian peculiar
acceleration in {\em comoving coordinates}. Therefore, we also need to
derive a relation between the proper acceleration $\vecg = \ddvecr$,
the proper peculiar acceleration $\vecg_{\rm pec}$, and the peculiar
acceleration in comoving coordinates $\vecF_{\rm pec}$. The second
derivative with respect to time of $\vecr = a \vecx$ gives

\begin{equation} \label{PecAcc}
 \ddvecr = a \ddvecx + 2\dot{a}\dot{\vecx} + \ddot{a} \vecx \ ,
\end{equation}

\noindent
whereas combing Eqs.~(\ref{eomPec}) leads to

\begin{equation}
 \ddvecx + 2 {\dot{a}\over a} \dvecx = {\vecF_{\rm pec}\over a^3}
\end{equation}

\noindent
Using the second Friedmann equation we can then rewrite \Eq{PecAcc} as
follows

\begin{equation}
 \ddvecr = {1\over a^2} \left( \vecF_{\rm pec} - {4\pi G \over 3} \overline{\rho} \vecx \right) \ .
\end{equation}

\noindent
This equation shows that the \textit{peculiar acceleration in proper
coordinates} $\vecg_{\rm pec}$ should be defined as

\begin{equation} \label{PecAcc2}
% \vecg_{\rm pec} = {\vecF_{\rm pec} \over a^2}
 \vecg_{\rm pec} = {\vecF_{\rm pec} / a^2}
\end{equation}

\noindent
where $\vecF_{\rm pec}$ is the solution of Poisson's equation
\Eq{PoissonPec} as obtained by \mlapm.

If we now make the assumption that MOND \textit{only} affects the
peculiar acceleration in proper coordinates but leaves the Hubble
acceleration unchanged, the recipe for getting the MONDian peculiar
accelerations in comoving coordinates $\vecF_{M, \rm pec}$ is as
follows:

\begin{enumerate}
 \item solve \Eq{PoissonPec} using \mlapm\ which gives $\vecF_{\rm pec}$
 \item use \Eq{PecAcc2} to transfer $\vecF_{\rm pec}$ to $\vecg_{\rm pec} = \vecF_{\rm pec}/a^2$
 \item use \Eq{ggN} to calculate $\vecg_{M, \rm pec}$ from $\vecg_{\rm pec}$
 \item transfer $\vecg_{M, \rm pec}$ back to $\vecF_{M, \rm pec} = a^2 \vecg_{M, \rm pec}$ 
 \item use $\vecF_{M, \rm pec}$ rather than $\vecF_{\rm pec}$ in \Eq{eomPec}
\end{enumerate}

This scheme has been employed for the simulations carried out with
\mlapm\ described below\footnote{The latest release version of \mlapm\
includes the MOND implementation which can be activated using
\texttt{-DMOND} on compile time.}. We like to point out that
our treatment of MOND agrees with the one advocated by Sanders (2001)
in the limit $\beta=0$.

%%%%%%%%%%%%%%%%%%%%%%%%%%%%%%%%%%%%%%%%%%%%%%%%%%%%%%%%%%%%%%%%%%%%%%%%
%                             MOND MODELS                              %
%%%%%%%%%%%%%%%%%%%%%%%%%%%%%%%%%%%%%%%%%%%%%%%%%%%%%%%%%%%%%%%%%%%%%%%%
\section{The Cosmological Models} \label{Models}
The reason for introducing MOND by Milgrom (1983) was to explain the
flat rotation curves of galaxies \textit{without} the need for dark
matter. Having this in mind we decided to use an $\Omega_0$ value for
the MOND simulation that is close to the upper bound allowed by
Big-Bang-Nucleosynthesis and agrees with the recent measurements of
cosmological parameters by the WMAP experiment (Spergel~\ea 2003). Our
database of simulations is made up of the following three runs

\begin{itemize}
 \item a standard \LCDM\ model,
 \item an open, low-$\Omega_0$ model with the same $\sigma_8$ as \LCDM,
 \item an open, low-$\Omega_0$ model with MOND and adjusted $\sigma_8$.
\end{itemize}

These runs are labeled \LCDM, OCBM and OCBMond, respectively, and
their cosmological parameters are summarized in
\Table{parameter}. The OCBM model is only to be understood as a gauge
for the MOND model run rather than an alternative to \LCDM.

We can see from \Eq{PecAcc2} that the peculiar accelerations (which
are subject to MOND) are large at early times and therefore a
"Newtonian treatment" in the early universe is justified.  Therefore,
the input power spectra to our initial conditions generator were
calculated using the CMBFAST code (Seljak~\& Zaldarriaga 1996) with
$\Omega_0 = \Omega_b = 0.04$ for OCBM and OCBMond, respectively. This
explains the choice for using the expression CBM rather than CDM. Such
a relatively high $\Omega_b$ value (compared to $\Omega_0$) actually
introduces "baryon wiggles" into the primordial power spectrum
(oscillations frozen into the plasma at the epoch of recombination
which are suppressed in dark matter dominated models). However,
fluctuations on scales of the box size employed ($B=32$\hMpc, see
below) and smaller are not affected by it other than an overall
"damping" (e.g. Eisenstein~\& Hu 1998; Silk 1968). This damping,
however, is compensated by the choice of normalisation $\sigma_8^{\rm
norm}$ of the power spectrum. We, moreover, differentiate between
$\sigma_8^{\rm norm}$ and the actual measure of $\sigma_8^{z=0}$ in
the simulations at redshift $z=0$ because the OCBMond model requires a
lower $P(k)$-normalisation $\sigma_8^{\rm norm}$ to arrive at a
comparable $\sigma_8^{z=0}$ value. This is due to a much faster growth
of structures when using MOND which will be emphasized in more detail
in Section~\ref{Analysis}.

\begin{table}
\caption{Model parameters. In all cases a value for the Hubble parameter
        of $h=0.7$ was employed.}
\label{parameter}
%\begin{center}
\begin{tabular}{lllllll}

label   & $\Omega_0$ & $\Omega_b$ & $\lambda_0$ & $\sigma_8^{z=0}$ & $\sigma_8^{\rm norm}$ & $g_0$ [cm/s$^2$]\\ 
\hline \hline
\LCDM\  &    0.30    &   0.04    & 0.7  &   0.88    & 0.88  & ---\\
OCBM    &    0.04    &   0.04    & 0.0  &   0.88    & 0.88  & ---\\
OCBMond &    0.04    &   0.04    & 0.0  &   0.92    & 0.40  & 1.2 $\times 10^8$ \\

\end{tabular}
%\end{center}
\end{table}

%%%%%%%%%%%%%%%%%%%%%%%%%%%%%%%%%%%%%%%%%%%%%%%%%%%%%%%%%%%%%%%%%%%%%%%%
%                         N-BODY SIMULATIONS                           %
%%%%%%%%%%%%%%%%%%%%%%%%%%%%%%%%%%%%%%%%%%%%%%%%%%%%%%%%%%%%%%%%%%%%%%%%
\section{The $N$-body Simulations} \label{Nbody}

Using the input power spectra derived with the CMBFAST code we
displace $128^3$ particles from their initial positions on a regular
lattice using the Zel'dovich approximation (Efstathiou, Frenk~\& White
1985). The box size was chosen to be 32\hMpc\ on a side. This choice
guarantees proper treatment of the fundamental mode which will still
be in the linear regime at $z=0$ (cf. the scale turning non-linear at
$z=0$ is roughly 20\hMpc\ for the models under investigation). The
particles were evolved from redshift $z=50$ until $z=0$ with the open
source adaptive mesh refinement code \mlapm\ (Knebe, Green~\& Binney
2001). We employed 500 steps on the domain grid built of $256^3$
cells, and in all three runs a force resolution of 11\hkpc\ was
reached in the highest density regions. The mass resolution of the
runs is $m_p = 1.30\cdot 10^{9}$\hMsun\ for the
\LCDM\ model and $m_p = 0.17\cdot 10^{9}$\hMsun\ for the two
low-$\Omega_0$ models, respectively. We output the particle positions
and velocities at redshifts $z=5,3,1,0.5$, and 0. These "snapshots"
are then analysed with respect to the large-scale clustering as well
as properties of individual objects. And even though in OCBM(ond) we
made the assumption of the non-existence of dark matter we still refer
to these objects as "halos"; due to the absence of dissipation they
show similar shapes and density distributions when being compared to
their \LCDM\ counterparts as can be seen in the analysis below.

Gravitationally bound objects are identified using the
Bound-Density-Maxima method (BDM, Klypin~\& Holtzman 1997).   The BDM code identifies local
overdensity peaks by smoothing the density field on a particular scale
of the order of the force resolution. These peaks are prospective halo
centres. For each of these halo centers we step out in
(logarithmically spaced) radial bins until the density reaches
$\rho_{\rm halo}(r_{\rm vir}) =
\Delta_{\rm vir} \rho_b$ where $\rho_b$ is the background density. 
This defines the outer radius $r_{\rm vir}$ of the halo. However, one
needs to carefully choose the correct virial overdensity $\Delta_{\rm
vir}$ which is much higher for the OCBM and OCBMond model due to the
low $\Omega_0$ value. The parameters used are $\Delta_{\rm vir}=340$
for \LCDM\ and $\Delta_{\rm vir}=2200$ for OCBM/OCBMond (see Gross 1997,
Appendix C, and references therein).

   \begin{figure*}
      \centerline{\psfig{file=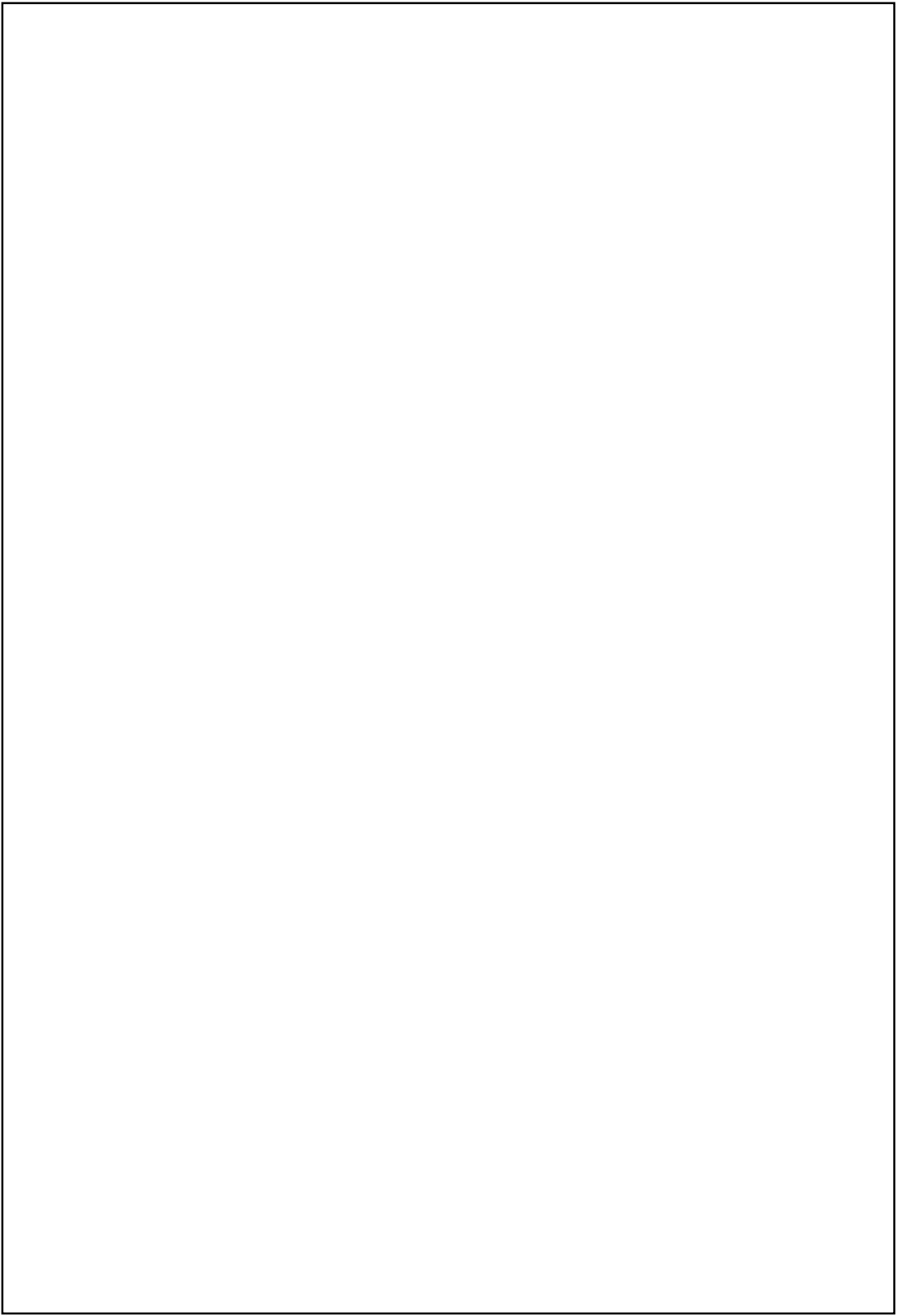,height=22cm}}
      \caption{Comparison of the large-scale density field of the
               three models under investigation at redshift $z=0$ (left)
               and $z=5$ (right). The upper panel
               shows the \LCDM\ simulation, the middle panel
               the OCBMond model and the lower one OCBM.} 
      \label{LCDMvsOCBMond}
   \end{figure*}

Brada~\& Milgrom (1999) raised the issue that care should be taken
when numerically integrating the equations of motion using MOND. Even
more so in our case where we made tha assumption that the curl-term in
\Eq{curl} vanishes which gives rise to forces that are not strictly
conservative. We therefore performed a simple test on the final output
to assure that our implementation of MOND does not violate momentum
conservation: the net momentum of all particles in the box as well as
the net force should vanish. We calculated the following sums

\begin{equation}
  \left|\sum_i \vec{v}_i\right| = C_1 \sum_i |\vec{v}_i| \ , \mbox{\ \rm and \ \ }
  \left|\sum_i \vec{F}_i\right| = C_2 \sum_i |\vec{F}_i| \ ,
\end{equation}

\noindent
and derived values for $C_1$ and $C_2$ in all three models. In
\LCDM\ and OCBM those constants $C_{1,2}$ are of the order $10^{-4}$ and even
though in OCBMond they are about an order of magnitude larger we still
believe that they are sufficiently small. It has been shown elsewhere
(Knebe, Green~\& Binney 2001) that adaptive softening in general does
not guarantee precise momentum conservation and our values for $C_1$
and $C_2$ are as expected.

%%%%%%%%%%%%%%%%%%%%%%%%%%%%%%%%%%%%%%%%%%%%%%%%%%%%%%%%%%%%%%%%%%%%%%%%
%                              RESULTS                                 %
%%%%%%%%%%%%%%%%%%%%%%%%%%%%%%%%%%%%%%%%%%%%%%%%%%%%%%%%%%%%%%%%%%%%%%%%
\section{Analysis} \label{Analysis}

%%%%%%%%%%%%%%%%%%%%%%%%%%
\subsection{Large-Scale Clustering Properties} 
\label{SECpower}
\begin{figure}
      \centerline{\psfig{file=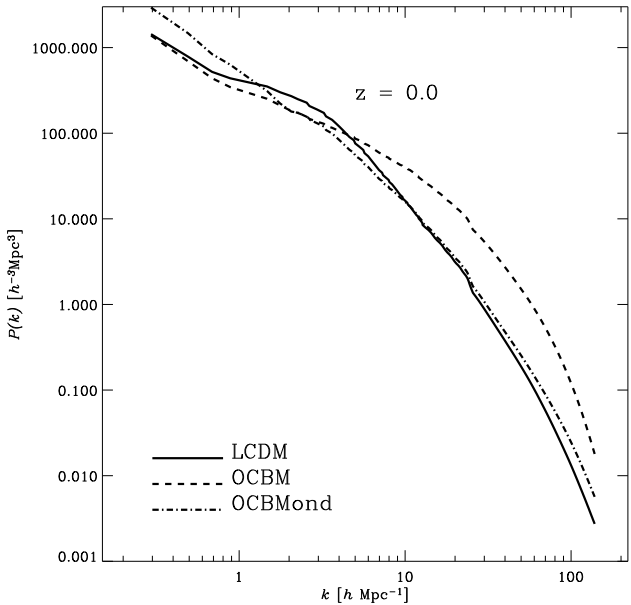,width=\hsize}}
      \caption{Matter power spectra for all three model 
               at redshift $z=0$ (thick lines) and $z=5$ (thin lines).}
      \label{power}
    \end{figure}

%%%%%%%%%%%%%%%%%%%%%%%%%%
We start with inspecting the large-scale density field in all three
runs. \Fig{LCDMvsOCBMond} shows a projection of the whole simulation
with each individual particle grey-scaled according to the local
density at redshift $z=0$ and $z=5$. This figure indicates that the
MOND simulation looks fairly similar to the other two models in terms
of the locations of high density peaks (dark areas), filaments and the
large-scale structure, respectively. One should bear in mind though
that the OCBMond simulation was started with a much lower
$\sigma_8^{\rm norm}$ normalisation than the other two runs. This is
in fact reflected in the right panel showing the density field at
redshift $z=5$; the MOND simulation is less evolved. Moreover, without
any further analysis one might even be inclined to conclude from
\Fig{LCDMvsOCBMond} that the MOND model is more strongly clustered at $z=0$
which is affirmed by the slightly higher $\sigma_8^{z=0}$ value given
in \Table{parameter}.  But as we will see later on this is not
necessarily true; we can confirm a higher amplitude of the two-point
correlation function for galaxies in OCBMond (cf. \Fig{halohaloXi})
while at the same time showing a comparable amplitude in the matter
power spectrum on small scales. The latter can be viewed in
\Fig{power} where we plot the matter power spectrum for all three
models at redshift $z=0$.  There we observe that the OCBMond model
shows a slightly larger amplitude for $k<0.8h{\rm Mpc}^{-1}$ (scales
close to the fundamental mode).  Nusser (2002), who treated the MOND
equation similarly (but differently) to us\footnote{Nusser (2002) did
not use Milgrom's interpolation function as given by
\Eq{MilgromInterpol} but rather a spontaneous transition from
Newtonian to MOND accelerations.}, already pointed out that the linear
evolution of the growing mode solution for the density contrast
$\delta$ scales like $\delta \propto a^2$ as opposed to Newtonian
theory where $\delta \propto a$. This explains why the OCBMond model
with the (initially) low $\sigma_8^{\rm norm}$ normalisation outruns
the evolution of the Newtonian OCBM simulation. In other words, the
MOND model had to be started with a lower $\sigma_8$ normalisation to
provide competitive results at redshift $z=0$.  Sanders (2001) also
noted that due to a much faster growth of structures in MOND universes
the amplitude of $P(k)$ for purely baryonic models matches the
standard \LCDM\ cosmology. We also like to point out that the required
value $\sigma_8^{\rm norm}=0.4$ needed to bring the MOND simulation
into agreement with the standard \LCDM\ model is closer to the COBE
normalisation $\sigma_{8}^{\tiny \textrm{COBE}} \approx 0.1$ for that
particular cosmology. Another feature worth noting is that the OCBMond
power spectrum does not show a distinctive "break" due to the transfer
of power from large to small scales as seen in both the
\LCDM\ and the OCBM model.

\begin{figure}
      \centerline{\psfig{file=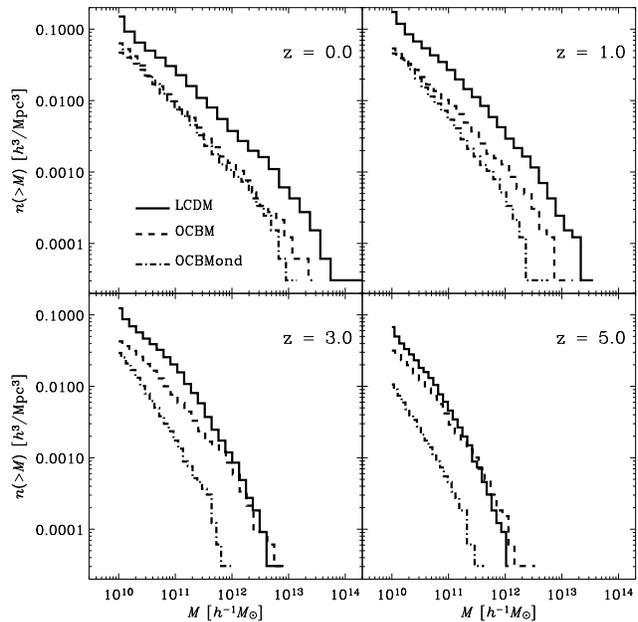,width=\hsize}}
      \caption{Cumulative mass function for all three models
               at redshifts $z=5,3,1$ and 0.}
      \label{massfunc}
    \end{figure}

\begin{figure}
      \centerline{\psfig{file=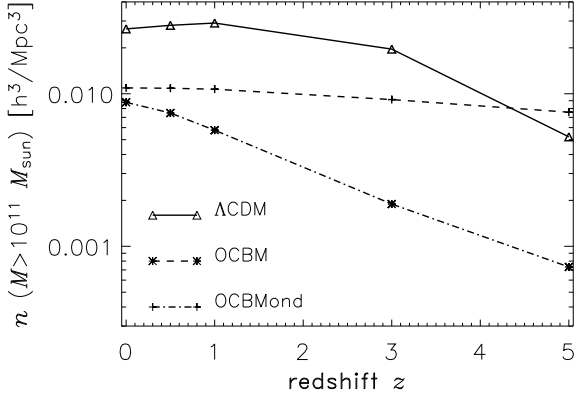,width=\hsize}}
      \caption{Redshift evolution of the abundance of halos 
               with mass $M>10^{11}$\hMsun.}
      \label{abundance}
    \end{figure}

\begin{figure}
      \centerline{\psfig{file=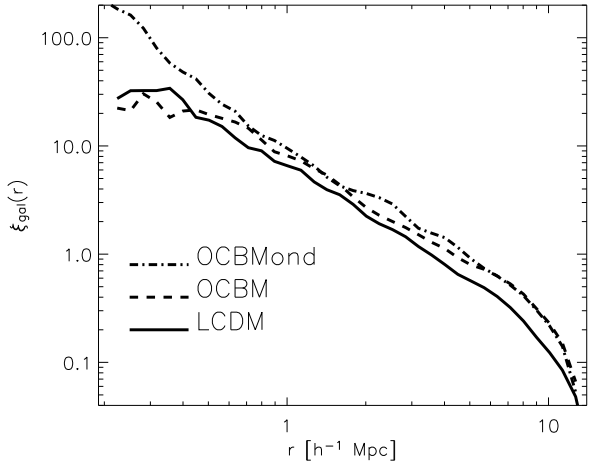,width=\hsize}}
      \caption{Two-point correlation function at redshift $z=0$ for
               the 500 most massive objects.}  
      \label{halohaloXi} 
   \end{figure}

In \Fig{massfunc} we are plotting the cumulative mass function of
objects identified using the BDM technique (Klypin~\& Holtzmann
1997). This figure highlights again that the hierarchical formation of
gravitationally bound objects is driven much faster in the MOND
simulation but being initiated at later times. At a redshift of $z=5$
we can see that the abundance of objects on all mass scales is nearly
identical in the OCBM and \LCDM\ model with a much lower amplitude for
OCBMond.  Whereas the evolution for the Newtonian OCBM run already
ceases at a redshift of around $z_{\rm stop} \simeq 1/\Omega_0 - 1
\simeq 24$ we still see a very strong increase (by more than one order
of magnitude) in the number density of objects of all masses in the
OCBMond simulation. To emphasize on this,
\Fig{abundance} shows the (integral) abundance evolution of objects
with mass $M>10^{11}$\hMsun.  The OCBM model undoubly experiences very
little evolution from $z\sim 5$ to today whereas both other models
show a very steep evolution. The discrepancy between the OCBMond and
the other two models at redshift $z=5$ can, however, be ascribed to
the lower initial $\sigma_8^{\rm norm}$ value though; OCBMond was set
up with much smaller initial density perturbation which only grew to a
comparable level of clustering via the effects of MOND. This is again
in agreement with the findings of Sanders (2001) who showed that the
collapse of spherically symmetric overdensities becomes MOND dominated
for redshifts $z\lesssim 5$ and hence starts to outrun Newtonian
models (cf. Fig.5 in Sanders 2001).

The question now arises to what degree the (formation) sites of halos
in \LCDM\ and the two OCBM models are correlated. To this extent we
calculated the two-point correlation function of the 500 most massive
objects and the result can be found in \Fig{halohaloXi}. We chose to
use a fixed number of halos rather than a mass cut for the calculation
of $\xi_{\rm gal}(r)$ in order not to introduce an artificial bias.
We do have far more objects of a given mass in the \LCDM\ model and
therefore a mass limit $M_{\rm min}$ used with the estimator for
$\xi_{\rm halo}(r)$ would lead to different correlation
amplitudes. The agreement between the Newtonian OCBM and the \LCDM\
model is not surprising. As already pointed out by other authors, the
correlation function is expected to be (nearly) identical in cases of
equal $\sigma_8$ normalizations, irrespective of the cosmological
model (Martel~\& Matzner 2000). Moreover, if the model is fixed and
only the $\sigma_8$ normalization varied it should leave no imprint on
$\xi_{\rm gal}(r)$ (Croft~\& Efstathiou 1994). But the OCBMond model
stands out again having a much higher amplitude on all scales.  This
clearly attributes for the differences already mentioned in the
discussion of \Fig{LCDMvsOCBMond} and indicates that the OCBMond model 
is more evolved at redhsift $z=0$ than the other two models even though 
the matter power spectrum has a lower amplitude on corresponding scales:
structure formation in MONDian cosmologies is even more biased than in
Newtonian models. This is in agreement with findings that small scales
enter the MOND regime before large-scale fluctuations (cf. Nusser
2002; Sanders 2001).

   \begin{figure}
      \centerline{\psfig{file=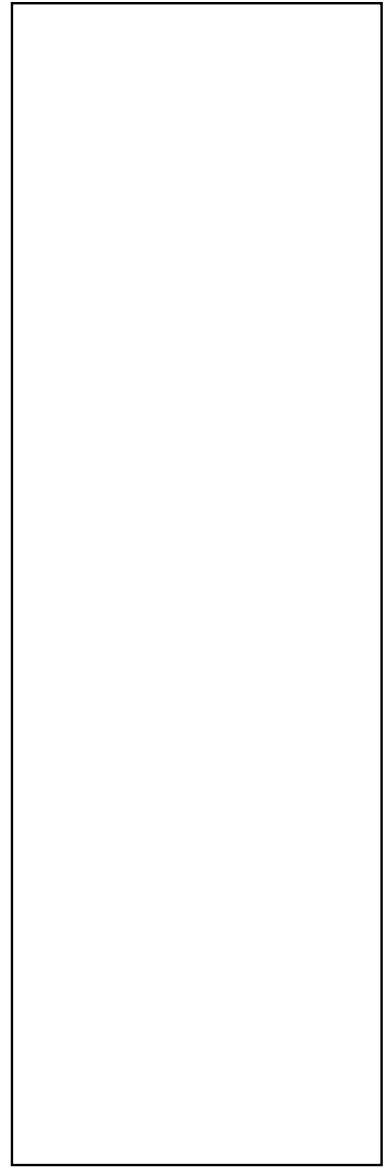,width=5cm}}
      \caption{Most massive galactic halo in \LCDM\ (upper panel),
               OCBMond (middle panel) and OCBM (lower panel). The 
               line in the lower right corner of each panel is a 
               scale indicating 100\hkpc.}
      \label{Haloes}
    \end{figure}

%%%%%%%%%%%%%%%%%%%%%%%%%%%%%
\subsection{Galactic Halos} 
%%%%%%%%%%%%%%%%%%%%%%%%%%%%%
Having analysed the large-scale clustering properties we now turn to
the investigation of the internal properties of galactic halos. To
this extent we use the halo catalogues based on the BDM code
(Klypin~\& Holtzmann 1997) and neglecting objects less massive than
$M<10^{11}$\hMsun\ again. This sets the minimum number of particles
per halo to 77 for \LCDM\ and 488 for OCBM(ond).

A visualization of the density fields throughout the most massive BDM
halo is given in \Fig{Haloes}. It is quite striking that neither of
the low-$\Omega_0$ halos shows substantial substructure. However, this
is easily understood for OCBM, because in a low-density universe
structure formation ceases at early times ($z_{\rm stop} \simeq
1/\Omega_0 - 1$).  This means that clusters in such cosmologies should
show fewer substructure since they had more time to virialize
(cf. Knebe~\& M\"uller 2000). But this explanation obviously does not
hold for OCBMond as nearly all halos in that particular model formed
exceptionally late (cf.~\Fig{abundance}).

\begin{figure}
      \centerline{\psfig{file=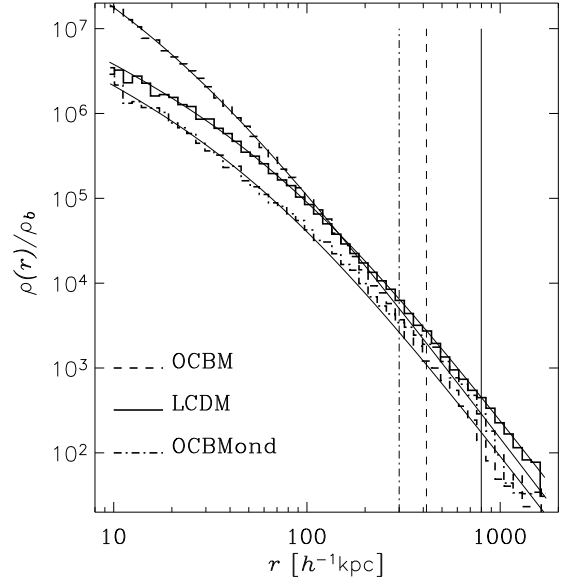,width=\hsize}}
      \caption{Density profile for the most massive halo in all
               three models. The vertical thin lines are indicating
               the virial radii.}
      \label{DensProfile}
    \end{figure}

The most interesting question, however, is probably the shape of the
density profile and the rotation curve for halos in MONDian
cosmologies, respectively. \Fig{DensProfile} now shows the matter
profile of the most massive halo in all three models along with fits
(thin solid lines) to a Navarro, Frenk~\& White (NFW) profile
(Navarro, Frenk~\& White 1997)

\begin{equation}\label{NFWfit}
 \rho_{\rm NFW}(r) \propto \frac{1}{r/r_s (1+r/r_s)^2} .
\end{equation}

\noindent
The scale radius $r_s$ is being used to define the concentration of
the halo $c=r_{\rm vir}/r_s$ where $r_{\rm vir}$ is the radius at
which the density reaches the virial overdensity $\Delta_{\rm vir}
\approx 340$ and $\Delta_{\rm vir} \approx 2200$ for \LCDM\ and
OCBM(ond), respectively.

\begin{figure}
      \centerline{\psfig{file=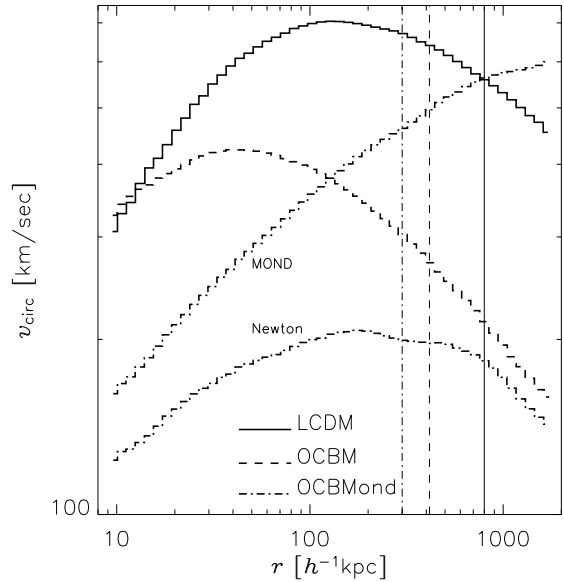,width=\hsize}}
      \caption{Rotation curve of the most massive halo.}
      \label{Vcirc}
    \end{figure}

We observe that even for the OCBMond model the data is equally well
described by the functional form of a NFW profile (out to the virial
radius, at least). However, the central density of that halo in the
OCBMond model is lower than in \LCDM\ and especially in OCBM.  The
high central density for OCBM is readily explained by the fact that
halos in that particular cosmology form at a time when the universe is
still very dense. This result is also supported by the values of the
concentration parameter presented in
\Table{DMhalo}: the most massive object in the MOND model shows the
lowest concentration $c$, mostly due to the late onset of formation as
observed in \Fig{abundance}. When inverting the density profile into a
(Newtonian) circular velocity curve by simply using $v_{\rm circ}^2(r)
= GM(\!<\!r)/r$, this also entails a shift of the maximum of the
rotation curve to higher radii as can be seen in
\Fig{Vcirc} ($r_{\rm max}^{\rm circ} \approx 2r_s$, Navarro,
Frenk~\& White 1997).  The functional shape of the rotation
curves for an NFW density profile is given by

\begin{equation} \label{Vmax}
 \frac{v^2_{\rm circ}}{v^2_{\rm vir}} = \frac{1}{x} 
                                        \frac{\ln{(1+cx)}-(cx)/(1+cx)}
                                             {\ln{(1+c)}-c/(1+c)} 
\end{equation}

\noindent
with $v_{\rm vir}$ being the (Newtonian) circular velocity at the virial radius
$r_{\rm vir}$ and $x=r/r_{\rm vir}$. The drop of the maximum of the
rotation curve by about a factor of 3.2 is merely a reflection of the
scatter in mass for the most massive halo throughout the three
models. As can be seen in \Table{DMhalo} the halo is more than ten
times as massive in \LCDM\ than in OCBMond. This should give an about
2.7 times higher $v_{\rm max}^{\rm circ}$ as the scaling between those
two quantities is roughly $v_{\rm max}^{\rm circ} \propto M^{1/3}_{\rm
vir}$. This scaling relation can be derived when using $x_{\rm
max}^{\rm circ} \approx 2/c$ (cf. Bullock~\ea 2001b and Navarro, Frenk~\& White
1997) together with \Eq{Vmax} giving

\begin{equation}\label{VcircMax}
 v_{\rm circ}^{\rm max} \propto v_{\rm vir} \sqrt{\frac{c}{\ln{(1+c)}-c/(1+c)}}
\end{equation}

\noindent
If we furthermore assume $c \propto M^{-0.13}$ as shown by Bullock~\ea
(2001b, cf. Eq(18)) we find that the $\sqrt{...}$-factor in
\Eq{VcircMax} is roughly constant for the mass range under
consideration.  And as $v_{\rm vir} \propto M^{1/3}$ (which
simply follows from $v^2 \propto M/r$ and $r \propto M^{1/3}$ for a 
spherical overdensity) the same scaling then holds for $v_{\rm
circ}^{\rm max}$ explaining the observed drop of the maximum of the rotation
curve for the OCBMond model.

However, for the OCBMond model we should take into account that
accelerations are \textit{not} Newtonian and hence $v_{\rm circ}^2(r)
= GM(\!<\!r)/r$ does not hold. Therefore we also show in \Fig{Vcirc}
the actual MONDian rotation curve calculated as follows: the Newtonian
acceleration $g = v^2/r$ is transfered to the MONDian acceleration
$g_M$ according to \Eq{ggN}. $g_M$ in turn is used to calculate
$v_{M}(r) = \sqrt{g_M r}$. The resulting $v_{\rm circ}(r)$ is labelled
"MOND" in \Fig{Vcirc}. We note that the MONDian velocities are
actually larger than the Newtonian ones bringing them closer to the
\LCDM\ model. This has implications for dynamical mass estimates of
galaxy clusters as presented in Sanders (1999). Sanders showed that
the dynamically estimated mass of galaxy clusters is too large
compared to the observed mass when using Newtonian physics. However,
MOND rectifies dynamical masses bringing them into better agreement
with observed masses. And a similar phenomenon can be observed in
\Fig{Vcirc}: the MONDian curve for OCBMond would be measured
observationally and hence be translated into a too high cluster mass
of $M_{\rm vir} \approx 2.8\cdot 10^{13}$\hMsun\ when using Newtonian
dynamics. MOND, on the contrary, would give the real value of $M_{\rm
vir}=0.3\cdot 10^{13}$\hMsun. Later on we will see that this leaves an
imprint on the (radial) distribution of the spin parameter, too.

\begin{figure}
      \centerline{\psfig{file=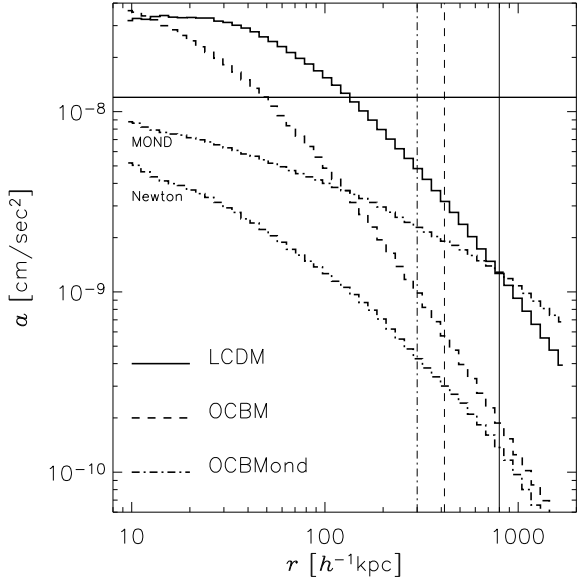,width=\hsize}}
      \caption{Acceleration curve of the most massive halo.
               The solid vertical line indicates the MONDian
               accleration parameter $g_0=1.2\times 10^8$cm/sec$^2$.}
      \label{a}
    \end{figure}

In addition to the rotation curve we also show the acceleration as a
function of radius in \Fig{a}. The object that formed in OCBMond has
MONDian accelerations throughout the whole halo whereas the central
parts of the objects in the other two models are above the universal
acceleration $g_0$ indicated by the thin solid line.

We like to stress that in both Figs~\ref{DensProfile} and \ref{Vcirc}
the Newtonian curves for the OCBMond model are only plotted for
completeness; they do \textit{not} carry observable information as the
halo actually follows MONDian physics rather than Newtonian. However,
they are educational in a sense to gauge the importance MOND has on
the internal structure of the halo.

\Table{DMhalo} now lists some internal properties in addition to the 
ones already mentioned, i.e. the velocity dispersion $\sigma_v$, the
virial radius $r_{\rm vir}$, the triaxiality $T$, ellipticities $e_1$
and $e_2$, the spin parameter $\lambda$ as well as best-fit parameters
$\lambda_0$ and $\sigma_{\lambda}$ when fitting the probability
distribution $P(\lambda)$ to a log-normal distribution. The spin
parameter $\lambda$ was calculated using the definition given in
Bullock~\ea (2001a)

\begin{equation} \label{lambda}
 \lambda = \frac{J}{\sqrt{2} M_{\rm vir} v_{\rm vir} r_{\rm vir}} \ ,
\end{equation}

\noindent
which proved to be a more stable measurement than the usual $\lambda'
= J \sqrt{|E|}/(G M^{5/2})$ definition. For the OCBMond model $v_{\rm
vir}$ is again the MONDian circular velocity at the virial radius. The
distribution of $\lambda$, $P(\lambda)$, has been fitted to a
log-normal distribution (e.g. Frenk~\ea 1988; Warren~\ea 1992; Cole~\&
Lacey 1996; Maller, Dekel~\& Somerville 2002; Gardner 2001)

\begin{equation} \label{lognormal}
 P(\lambda) = \displaystyle \frac{1}{\lambda \sqrt{2\pi} \ \sigma_0}
              \exp \left( {-\frac{\ln^2 (\lambda/\lambda_0)}{2 \sigma_0^2}} \right) \ .
\end{equation}

\begin{figure}
      \centerline{\psfig{file=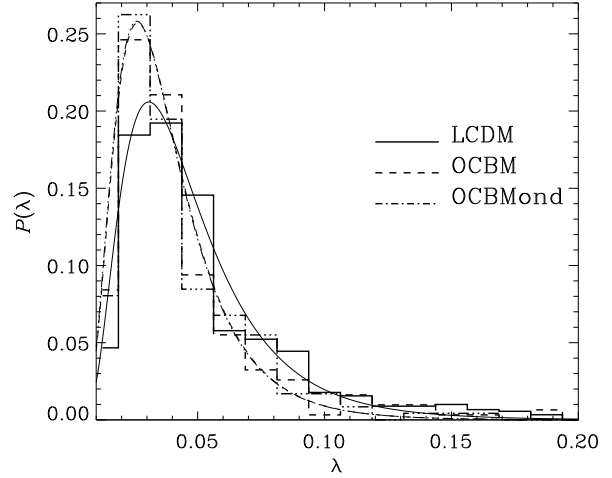,width=\hsize}}
      \caption{Spin parameter distribution in all three models.  Lines
      show fits obtained using the log-normal distribution given by
      Eq.~(\ref{lognormal}).}  \label{SpinFit} \end{figure}

\noindent
The results are presented in \Fig{SpinFit} showing that the OCBMond
distribution $P(\lambda)$ is nearly indistinguishable from the other
two models. However, the reduced $\chi^2$ value for OCBMond is about a
factor of two larger (cf. \Table{DMhalo}). As been noted by Maller,
Dekel~\& Sommerville (2001) the log-normal distribution given
by~\Eq{lognormal} is not as good a fit to models with halos that are
recent merger remnants. This is definitely one of the effects that has
an influence on the spin parameter distribution for the OCBMond model
as we expect a high level of recent merger activity
(cf. \Fig{abundance}).

\begin{figure}
      \centerline{\psfig{file=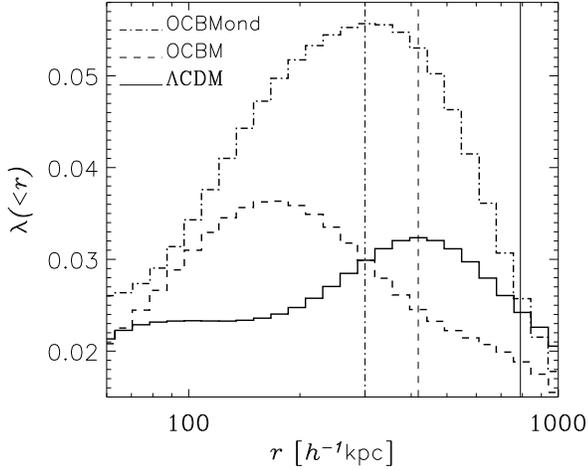,width=\hsize}}
      \caption{Radial distribution of spin parameter $\lambda$.
               The vertical lines are again showing the virial radius.}
      \label{SpinProfile}
    \end{figure}

We also calculated the radial distribution of $\lambda (<\!r)$
throughout the halos and present the result for the most massive one
in \Fig{SpinProfile}. We see that $\lambda(<\!r)$ is roughly constant
for the Newtonian models of structure formation whereas there is a
sharp increase of $\lambda(<\!r)$ in the MOND halo towards the virial
radius. This implies that the material in that particular halo moves
on more circular orbits and the object itself is closer to solid body
rotation, respectively. This result is in agreement with our previous
finding where we showed that the MONDian rotation curve is not given
by simply "inverting the density profile" as in the Newtonian case
(cf. \Fig{a}); the velocity in the outskirts of the halo is much
larger and hence we expect a rise in $\lambda(r)$, too.

We are now going to focus on the shape of the halos. Firstly, we show
measurements of the overall shape of halos and to this extent we
calculated the eigenvalues $a>b>c$ of the inertia tensor. They were
in turn used to construct the triaxiality parameter (e.g. Franx,
Illingworth~\& de Zeeuw 1991):

\begin{equation}
T=\frac{a^2-b^2}{a^2-c^2} \ .
\end{equation}

\noindent
The triaxialities $T$ are accompanied by the ellipticities

\begin{equation}
e_1 = 1-\frac{c}{a}, \ \ e_2 = 1-\frac{b}{a} 
\end{equation}

\noindent
and the values for the most massive object are summarized in
\Table{DMhalo}. The mean values $\langle T \rangle$, $\langle e_1
\rangle$, and $\langle e_2\rangle$ when averaging over all halos 
more massive than 10$^{11}$\hMsun\ can be found in \Table{shapes}. We
observe a trend for the MOND halos to be more triaxial with higher
ellipticities at the same time.

\begin{table*}
\caption{Properties of the most massive gravitionally bound object.
         The mass $M$ is measured in \hMsun, velocities in km/sec and
         radii in \hkpc. $\chi^2$ is actually
         the reduced $\chi^2$ value as returned by IDL's \texttt{curvefit} routine.}
\label{DMhalo}
%\begin{center}
\begin{tabular}{lccccccccccccc}
 model   &  $M$         
         &  $\sigma_v$
         &  $v_{\rm vir}$
         &  $r_{\rm vir}$
         &  $r_s$
         &  $c$ 
         &  $T$
         &  $e_1$
         &  $e_2$
         &  $\lambda$ 
         &  $\lambda_0$
         &  $\sigma_{\lambda}$ 
         &  $\chi^2$ \\ \hline \hline

\LCDM   & 5.6$\cdot 10^{13}$ & 657 & 553 & 790 & 91  & 8.7  & 0.65 & 0.30 & 0.18 & 0.025 & 0.040 & 0.53 & 0.8 \\
OCBMond & 0.3$\cdot 10^{13}$ & 286 & 473 & 300 & 59  & 5.1  & 0.81 & 0.30 & 0.23 & 0.057 & 0.034 & 0.52 & 1.5 \\
OCBM    & 0.8$\cdot 10^{13}$ & 373 & 271 & 418 & 27  & 15.5 & 0.80 & 0.21 & 0.16 & 0.023 & 0.034 & 0.52 & 0.9 \\
\end{tabular}
%\end{center}
\end{table*}

\begin{table}
\caption{Mean triaxiality parameter $\langle T\rangle $ and mean ellipticities 
         $\langle e_1\rangle $, $\langle e_2\rangle $ when averaging over
         halos with $M>10^{11}$\hMsun.}
\label{shapes}
%\begin{center}
\begin{tabular}{lccc}

label   & $\langle T\rangle$ & $\langle e_1\rangle$ & $\langle e_2\rangle$ \\
\hline \hline

\LCDM\  & 0.55  & 0.25    & 0.13\\
OCBMond & 0.61  & 0.28    & 0.17\\
OCBM    & 0.55  & 0.21    & 0.12\\

\end{tabular}
%\end{center}
\end{table}

\begin{figure}
      \centerline{\psfig{file=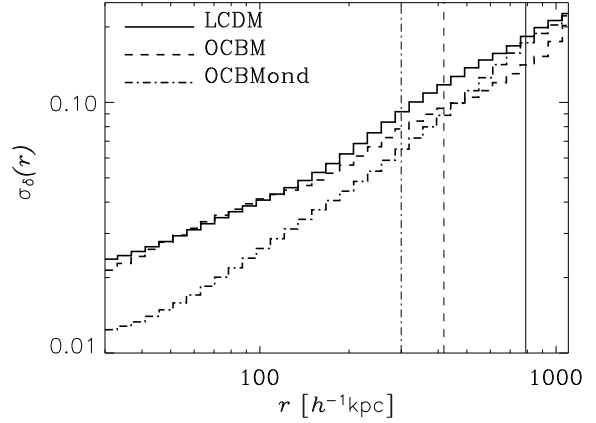,width=\hsize}}
      \caption{Variance $\sigma_{\delta}(r)$ for the most
               massive halo. The vertical lines are indicating
               the virial radius of the respective halo.}
      \label{sigD}
    \end{figure}

Secondly, we like to quantify subclumps within the virial radius of
the halo itself.  One possibility to measure the substructure content
of a halo is to calculate the radial profile of the density dispersion

\begin{equation} \label{sigmaD}
 \sigma^2_{\delta}(r) = \frac{1}{N(r)} \sum_{i=1}^{N(r)}
                        \left( \frac{\rho_i(r) - \langle \rho(r) \rangle}
                                  {\langle \rho(r) \rangle} \right)^2 \ ,
\end{equation}

\noindent
where $\rho_i(r)$ is the local density at a particle position which
was estimated using the nearest 20 neighbors and $\langle \rho(r)
\rangle$ the average taken over all $N(r)$ particles within a spherical 
shell $[r,r+dr]$. We used the same binning as already applied to the
density profile and the rotation curve, respectively. The result for
the most massive halo in all three runs is presented in
\Fig{sigD}. This figure shows that the dispersion  
is always smallest for the MOND halo, at least out to the virial
radius: the density at each individual particle position is always
close to the mean density. If there were subclumps present one would
expect to find peaks in the $\sigma^2_{\delta}(r)$ curve due to local
deviations from the mean density profile.

%%%%%%%%%%%%%%%%%%%%%%%%%%%%%%%%%%%%%%%%%%%%%%%%%%%%%%%%%%%%%%%%%%%%%%%%
%                          CONCLUSIONS                                 %
%%%%%%%%%%%%%%%%%%%%%%%%%%%%%%%%%%%%%%%%%%%%%%%%%%%%%%%%%%%%%%%%%%%%%%%%
\section{Summary and Discussion} \label{Conclusions}
In this paper we presented three cosmological simulations: a fiducial
standard \LCDM\ model, a very low-$\Omega_0$ model, and the same
$\Omega_0$ model but with MONDian equations of motions.  Putting aside
the classical arguments against MOND, we showed that structures found
in a MONDian low-$\Omega_0$ universe are not significantly different
from the standard \LCDM\ model. However, we derived several
differences which might easily validate or rule out MONDian
cosmologies observationally.  Our main results can be summarized as
follows:

\begin{itemize}
 \item even though the OCBMond run was set up using a low value for
       $\sigma_8^{\rm norm}=0.4$ the simulation
       finished with $\sigma_8^{z=0}=0.92$ (which is close to the
       cluster normalisation of the \LCDM\ model),

 \item the OCBMond model shows an extremely fast evolution of the number density
       of halos for redshifts $z<5$,

 \item virialized objects in a MONDian universe are slightly more correlated,

 \item the density profile of MOND objects still follows the universal
       NFW density profile, but they are 

 \item far less concentrated,

 \item show less substructure, and

 \item are closer to solid body rotation.

We conclude that the most distinctive feature of a MONDian universe is
the late epoch of galaxy formation; in a MONDian universe that is
normalised to match a \LCDM\ model at redshift $z=0$ we expect to
\textit{not} observe any galaxies until recently ($z<5$).  This
actually holds for any low-$\sigma_8$ universe either MONDian or
Newtonian, but only the assumption of MOND can bring such a model into
agreement with observations of the local universe again.  Another
interesting finding is that the outer parts of MOND halos are closer
to solid body rotation than their standard \LCDM\ counterparts even
though the overall distribution $P(\lambda)$ is nearly
indistinguishable from the
\LCDM\ model. 

However, there have still been many assumptions made during the course
of this study which are hard to justify and hence all results are to
be understood simply as preliminary until our understanding of MOND
actually improves. We not only assumed that MOND only affects peculiar
accelerations in proper coordinates but also neglected the curl-term
in \Eq{curl}. The effect of this rotational component is not clear,
but it is guaranteed to decrease rapidly on large scales and vanish
completely for objects that have spherical, planar or cylindrical
symmetry. Another disclaimer is that MOND is based on the
non-existence of dark matter, but our simulations only model gravity;
if the universe consist only of baryons one definitely would need to
include gas physics to be able to make credible predictions for
internal properties of galaxies. But this is beyond the scope of this
study. Nonetheless, we have shown that cosmology with MOND does not
necessarily lead to completely odd results. Our treament of MOND
though gives clustering properties comparable to the favorite
concordance \LCDM\ model.

\end{itemize}

%%%%%%%%%%%%%%%%%%%%%%%%%%%%%%%%%%%%%%%%%%%%%%%%%%%%%%%%%%%%%%%%%%%%%%%%
%                         ACKNOWLEDGEMENTS                             %
%%%%%%%%%%%%%%%%%%%%%%%%%%%%%%%%%%%%%%%%%%%%%%%%%%%%%%%%%%%%%%%%%%%%%%%%
\section*{Acknowledgments}
The simulations presented in this paper were carried out on the
Beowulf cluster at the Centre for Astrophysics \& Supercomputing,
Swinburne University. We acknowledge the support of the Australian
Research Council through its Discovery Project program (DP0343508).
We would like to thank the two anonymous referees for valuable
comments that helped to improve the scientific content of the paper.

%%%%%%%%%%%%%%%%%%%%%%%%%%%%%%%%%%%%%%%%%%%%%%%%%%%%%%%%%%%%%%%%%%%%%%%%
%                           BIBLIOGRAPHY                               %
%%%%%%%%%%%%%%%%%%%%%%%%%%%%%%%%%%%%%%%%%%%%%%%%%%%%%%%%%%%%%%%%%%%%%%%%
%========================

\end{document}